\documentclass[aps,prb,reprint,amsmath,amssymb,groupedaddress]{revtex4-1}
\usepackage{graphicx}% Include figure files
\usepackage{dcolumn}% Align table columns on decimal point
\usepackage{bm}% bold math
\usepackage{color}
\usepackage{amsmath}

\begin{document}

%\preprint{}

\title{DFT+DMFT study of spin-charge-lattice coupling in covalent LaCoO$_3$}


\author{Hyowon Park$^{1,2}$, Ravindra Nanguneri$^{1,3}$, Anh T. Ngo$^{2}$}
%\thanks{}
\affiliation{$^1$Department of Physics, University of Illinois at Chicago, Chicago, IL 60607, USA \\
$^2$Materials Science Division, Argonne National Laboratory, Argonne, IL, 60439, USA \\
$^3$Department of Chemistry, Brown University, Providence, RI 02912, USA}


\date{\today}

\begin{abstract}
We study energetics and the nature of both homogeneous and mixed spin (MS) states in LaCoO$_3$ incorporating structural changes of the crystal volume expansion and the Co-O bond disproportionation (BD) during the spin-state transition using the density functional theory plus dynamical mean field theory (DFT+DMFT) method.
DFT+DMFT predicts that energetics of both excited spin states are almost the same while DFT+U calculations of the same structures energetically favor the MS states and produce various metastable solutions whose energetics depend sensitively on final spin states.
%Within DFT+DMFT, the multi-configurational nature of excited spin-states exhibits smooth crossover from the low spin state during the spin-state transition and both excited spin-states are energetically almost the same while the DFT+U calculation of the same structure produces various meta-stable solutions with energetics depending sensitively on spin-states.
Within DFT+DMFT, the homogeneous spin state in the expanded crystal volume shows the multiconfigurational nature with non-negligible occupancy probabilities of both high spin (HS) and low spin (LS) states along with $d^6$ and $d^7$ charge configurations indicating the dynamically fluctuating nature of spin and charge states due to the Co-O covalency.
The nature of the MS state under the BD structure reveals that Co sites with the long Co-O bonds develop a Mott insulating state and favor HS with a $d^6$ configuration, while more covalent Co sites with the short Co-O bonds occupy more LS states with a $d^7$ configuration and behave as a band insulator, as a result, charge ordering is induced in the BD structure from the spin-state ordering. 
We also find that both energetics and electronic structure sensitively depend on the Co-O covalency effect, which can be tuned by changing the double counting potential and the resulting $d-$occupancy ($N_d$), and $N_d$ close to 6.7 is consistent with the nature of the spin-state transition. 
Our results show that structural changes during the spin-state transition can play an important role in understanding energetics and electronic structure of LaCoO$_3$.  
%The effects of the charge-self-consistency within DFT+DMFT on the results are also discussed. 
%
\end{abstract}

%\keywords{}

\maketitle

\section{Introduction}
\label{sec:intro}

Transition metal oxides exhibit complex and rich phase diagrams arising from the strongly correlated nature of spin, charge, orbital, and lattice degrees of freedom~\cite{RevModPhys.70.1039}.
%One of the fascinating facets in these materials is that the spin, charge, and lattice degrees of freedom are strongly coupled to each other and these competing electronic orders can be easily pinned by subtle structural modifications.
LaCoO$_3$ has been known for the spin-state transition of partially filled $d$ orbitals in a Co ion.
At very low temperatures, LaCoO$_3$ is a non-magnetic insulator with the low spin (LS) state. As the temperature is elevated above 90K,
the magnetic susceptibility changes to a Curie-Weiss form indicating that paramagnetism dominates with higher spin states while retaining an insulating behavior~\cite{Ivanova_2009}.
The spin-state transition can be explained based on the atomic multiplet structure of the Co $d$ orbital, namely from the $|S_z|$=0 LS ground state to $|S_z|$=1 intermediate spin (IS) or $|S_z|$=2 high spin (HS) state.
% based on the atomic multiplet structure of the Co $d$ orbital.
Various experimental results have been used to interpret the higher spin state as either
IS~\cite{PhysRevB.55.4257,PhysRevB.47.16124,PhysRevLett.99.047203,PhysRevB.66.020402} or HS~\cite{PhysRevLett.97.176405,PhysRevLett.97.247208}.
The MS of LS and HS has been also suggested to explain other experimental measurements~\cite{PhysRev.155.932,PhysRevB.5.4466,SENARISRODRIGUEZ1995224,PhysRevB.71.024418,PhysRevB.93.155137}.
Despite extensive experimental works, the nature of excited spin-states has not been clarified yet.
%Despite extensive experimental and theoretical efforts, the nature of excited spin states and the interplay between electron and lattice degrees of freedom has not been clarified yet.

The spin-state transition in LaCoO$_3$ occurs since the Hund's coupling tends to maximize the spin and excite electrons from $t_{2g}$ to $e_g$ orbitals by overcoming the crystal field splitting between them. 
Here, the interplay between electron and lattice degrees of freedom plays an important role as the $e_g$ orbital occupation increases the Co-O bond-length to reduce the repulsive Coulomb interaction of electrons between Co and O ions. 
This structural change also gives the positive feedback since the reduction of the crystal-field splitting can favor the spin-state transition.
%
%ransition occurs as the the Hund's coupling driven by electronic correlations can due to the balance between 
%While the spin-state transition can be driven by electronic correlations due to the Hund's coupling, the coupling between electron and lattice degrees of freedom is also expected since higher spin states usually accompany the CoO$_6$ octahedral volume expansion. This is because the increase of the $e_g$ orbital occupancy enhance the repulsive Coulomb interaction between Co and O electrons and then the volume expansion can reduce energetically this interaction. This structural change can also give the positive feedback to the spin-state transition since the crystal-field splitting is reduced and the higher spin state can be promoted more efficiently.
This strong electron-lattice coupling has been measured experimentally by the anomalous lattice expansion arising due to the Co-O bond-length elongation at the spin-state transition~\cite{PhysRevB.50.3025,PhysRevB.78.134402,EPJB.2005,PhysRevB.66.094408}.
Several scenarios of local structural distortions due to the spin excitation have been proposed although clear experimental evidences have not been given yet.
The Co-O bond-disproportionation (BD) with alternating the long bond (LB) site and the short bond (SB) site~\cite{PhysRev.155.932,PhysRevB.69.134409} was suggested to accommodate MS with HS and LS.
%The possibility of the excitation to mixed spins with HS and LS has also suggested the Co-O bond-disproportionation (BD) with alternating the long bond (LB) site and the short bond (SB) site~\cite{PhysRev.155.932,PhysRevB.69.134409}.
The Jahn-Teller distortion~\cite{PhysRevB.66.020402,LTP.32.162} was also discussed possibly due to the IS state.
% although this distortion is not compatible with the crystal symmetry.
%In addition to the structural distortion due to the spin-state transition,
The strong electron-lattice coupling has been also shown in the tensile-strained LaCoO$_3$ film promoting various competing orders including spin~\cite{PhysRevB.75.144402,APL.93.212501,nanoletter.12.4966}, charge~\cite{PhysRevLett.120.197201}, and orbital orderings~\cite{PhysRevLett.111.027206}.
%It has been also shown that structural modification of
%it has been shown that various forms of competing electronic orders can be pinned by structural modifications.
%The strong electron-lattice coupling of LaCoO$_3$ also suggests that electronic orders can be pinned by structural modifications.
%For example, LaCoO$_3$ modified to the surface state~\cite{PhysRevB.70.014402} or nanoparticles~\cite{PhysRevB.76.172407} can induce long-range ferromagnetism.
%For example, tensile-strained LaCoO$_3$ films also promotes various competing orders including spin~\cite{PhysRevB.75.144402,APL.93.212501,nanoletter.12.4966}, charge~\cite{PhysRevLett.120.197201}, and orbital ordering~\cite{PhysRevLett.111.027206}.

Alongside experimental measurements, various theoretical scenarios based on first-principle calculations have been proposed to address this long-standing problem of the spin-state transition in LaCoO$_3$. 
Density functional theory (DFT)+U calculations have
been predicting that excited spin-states including IS~\cite{PhysRevB.54.5309,PhysRevB.71.054420} and MS~\cite{Kn_ek_2006,PhysRevB.79.014430} states can be energetically stable.
%However, the strong correlation effect in DFT+U is treated in a static limit and accompanied with a long-range spin ordering, in which the spin symmetry must be broken explicitly during the calculation.
Dynamical mean field theory (DMFT) can capture the multi-configurational nature of a paramagnetic state fluctuating dynamically beyond DFT+U~\cite{RevModPhys.68.13,RevModPhys.78.865}.
Early DFT+DMFT studies computed the single-particle spectra with spin-state crossover~\cite{PhysRevB.77.045130} comparable to the experimental X-ray absorption spectra~\cite{PhysRevB.47.16124} and also studied effects of the pressure~\cite{PhysRevB.86.184413} and the Co-O covalency~\cite{PhysRevB.86.195104} on the spin-state transition.
%Effects of the pressure~\cite{PhysRevB.86.184413} and the Co-O covalency~\cite{PhysRevB.86.195104} on the nature of the spin-state transition also have been studied.
Both the homogeneous spin excitation including the electronic entropy~\cite{PhysRevMaterials.1.064403} and the mixed LS and HS solution without any structural distortions~\cite{PhysRevLett.106.256401,PhysRevLett.115.046401} have been discussed as the possible origin  of the spin-state transition within DFT+DMFT.
%The mixed LS and HS solution was also obtained within DFT+DMFT without any structural with a possible charge ordering has been obtained within DFT+DMFT from the pure electronic origin without the need of structural BD~\cite{PhysRevLett.106.256401,PhysRevLett.115.046401}.
%Recently, charge-self-consistent (CSC) DFT+DMFT has been used to elucidate roles of the electronic entropy and the Co-O octahedral rotation~\cite{PhysRevMaterials.1.064403}.
However, no DFT+DMFT studies have addressed yet both the nature and energetics of paramagnetic states with all possible structural changes in this material.
%However, no DFT+DMFT studies have addressed yet the effect of lattice changes on both homogeneous and mixed spin and charge states in LaCoO$_3$ occuring during the spin-state transition. 
%yet the interplay between the nature of both homogeneous and mixed spin states incorporating possible lattice responses by computing the energetics of possible spin-state transitions.

In this paper, we adopt the DFT+DMFT method to study both homogeneous and MS states by incorporating structural changes during the spin-state transition. 
We show that the DFT+DMFT energetics treating the multi-configurational nature of a paramagnetic state can be noticeably different from the static DFT+U solutions in which various meta-stable spin-states are possible.
%We will show that the multi-configurational nature of paramagnetic state treated in DFT+DMFT produces very close energetics of both spin states with the smooth crossover from LS states while DFT+U can have various meta-stable solutions.
%We will show that the nature and energetics of the excited spin state treated in DFT+DMFT is noticeably different from the solution of DFT+U.
We also find that both structural changes and the Co-O covalency effect tuned by the double counting potential can strongly affect the electronic structure and energetics of spin-states in LaCoO$_3$.
In the expanded crystal volume, the occupation probabilities of higher spin-states in the Co ion increase as Co $d$ orbitals become more correlated.
As temperature increases, DFT+DMFT can reproduce the insulator-to-metal transition, consistent with experiment, as the correlation in the Co ion becomes weaker.
In the BD structure, the Co ion in the Co-O LB behaves as a Mott insulator with HS while the Co ion in the SB becomes a band insulator favoring LS.
The charge ordering is also induced in the BD structure as the LB Co ion favors a $d^6$ configuration while the SB Co ion occupies more $d^7$ states.
The charge-self-consistency in DFT+DMFT plays a role to reduce the charge ordering in the BD structure.

This paper is organized as follows.
In Sec.$\:$\ref{sec:method}, the method used in this paper is explained in details.
Sec.$\:$\ref{sec:result1} discusses the energetics of possible spin-state transition in LaCoO$_3$ comparing DFT+DMFT and DFT+U.
The nature of the paramagnetic state treated in DFT+DMFT is shown in Sec.$\:$\ref{sec:result2} by computing the occupation probabilities of different spin-states. The origin of charge ordering driven by spin-state ordering in the BD structure is revealed in Sec.$\:$\ref{sec:result3}.
Results of the density of states computed using DFT and DFT+DMFT for different structures and temperatures are shown in Sec.$\:$\ref{sec:result4}. 
The self-energy data are displayed in Sec.$\:$\ref{sec:result5} and the effect of the charge-self-consistency in DFT+DMFT is also discussed in Sec.$\:$\ref{sec:result6}.
We summarize our paper with conclusions in Sec.$\:$\ref{sec:conclusion}.
%interplay among spin, charge, and lattice degrees of freedom in LaCoO$_3$.
%We find that the Co-O covalency effect can strongly affect both the energetics and the nature of spin-states accompanied by possible lattice responses.

%To study the interplay between the spin-state transition and possible structural changes in LaCoO$_3$,
%we first perform the structural relaxation
%%to study possible structural changes in LaCoO$_3$ due to spin-state transitions.
%by adopting the DFT+U method as implemented in the Vienna ab-initio simulation package (VASP)~\cite{vasp_1,vasp_2,vasp_3,vasp_4} code.
%The Perdew-Burke-Ernzerhof (PBE) functional is used for the exchange-correlation functional within DFT.
%The DFT+U convergence is achieved using the plane-wave energy cut-off of 600eV and the $k-$point mesh of 8$\times$8$\times$8. The convergence of the structural relaxation is achieved if the atomic forces of all ions are smaller than 0.01eV/\AA.
%Within DFT+U, we use the on-site Hubbard interaction U=6eV, which is obtained using the constraint DFT method~\cite{PhysRevB.86.195104}, and the Hund's coupling J=0.9eV.

\section{Method}
\label{sec:method}

\subsection{Structural relaxation}

%owever, our modified $V^{DC}$ form is still different
To study possible structural distortions during the spin-state transition,
% incorporating possible structural changes in LaCoO$_3$,
we first perform the structural relaxation by adopting the DFT+U method as implemented in the Vienna ab-initio simulation package (VASP) code~\cite{vasp_1,vasp_2,vasp_3,vasp_4} using different spin-states as initial guess.
The Perdew-Burke-Ernzerhof (PBE) functional~\cite{PBE} is used for the exchange-correlation functional within DFT.
The DFT+U convergence is achieved using the plane-wave energy cut-off of 600eV and the $k-$point mesh of 8$\times$8$\times$8. The convergence of the structural relaxation is achieved if the atomic forces of all ions are smaller than 0.01eV/\AA.
Within DFT+U, we use the on-site Hubbard interaction U=6eV, which is obtained using the constraint DFT method~\cite{PhysRevB.86.195104}, and the Hund's coupling J=0.9eV.
Although the ground-state of LaCoO$_3$ is paramagnetic, DFT+U relaxation calculations are performed  with the ferromagnetic configuration since correlations are included by imposing a long-range magnetic order in DFT+U.
%Different spin states can be imposed in the beginning of the relaxation to allow structural  

LaCoO$_3$ is a rhombohedral structure with the $R\bar{3}c$ symmetry containing two Co ions per unit cell.
The experimental crystal volume $V$ is 56.0\AA$^3$ per formula unit at low temperatures~\cite{PhysRevB.66.094408}.
%The experimental structure of LaCoO$_3$ is rhombohedral with the $R\bar{3}c$ symmetry with two Co ions per unit cell with the crystal volume $V$ of 56.0\AA$^3$ per formula unit at low temperatures~\cite{PhysRevB.66.094408}.
We find that the DFT+U relaxation calculation with the PBE functional converging to LS produces the volume $V$ of 56.40\AA$^3$ per formula unit with the Co-O bond-length $a\sim1.95$\AA$\:$ (S1 structure in Table.$\:$\ref{tab:relax}),
%We find that DFT+U relaxation calculations are converged to various meta-stable solutions depending on initial spin states which can be obtained by explicitly breaking spin and orbital symmetries
%The final spin-state does not change much from the initial configuration.
%The structural relaxation performed using the LS state produces the volume $V$ of 112.80\AA$^3$ with the Co-O bond length $a$ of 1.95\AA
while the IS state (magnetic moment=2.2$\mu B$) results in the $2.7$\% volume expansion resulting $V\sim57.98$\AA$^3$ per formula unit and $a\sim1.97$\AA (S2 structure in Table.$\:$\ref{tab:relax}).
The HS structure converged to a more expanded volume but the total energy is much higher than either LS or IS one, therefore we do not consider the HS structure in this paper.
%The relaxation with the IS configuration results in the $\sim$2.7\% volume expansion compared the LS volume producing $V\sim$115.96\AA$^3$ with the Co-O bond-length elongation to $d\sim$1.97\AA while the HS structure converged to an even more expanded volume with an even higher energy than either LS or IS energies.
Interestingly, the MS imposing HS to one Co ion and LS to the other Co ion within the unit-cell produces the BD structure by lowering the crystal symmetry from $R\bar{3}c$ to $R\bar{3}$ (S3 structure in Table.$\:$\ref{tab:relax}).
The crystal volume $V$ is similar to the IS volume ($V\sim57.98$\AA$^3$) and the HS site (magnetic moment=3$\mu B$) becomes a Co-O LB site with $a\sim1.99$\AA$\:$ and the LS site (magnetic moment=0.3$\mu B$) is the SB with $a\sim1.94$\AA, resulting in the bond-length difference $\delta a\sim0.05$\AA.
The $d-$occupancy ($N_d$) for relaxation results in all structures are close to $d^7$ (leaving one hole in surrounding oxygen atoms) although the nominal $d-$occupancy is $d^6$ for LaCoO$_3$.
This is due to the strongly covalent nature of the Co-O bonding in LaCoO$_3$.
%We also relax the structure with an initial MS state imposing HS to one Co ion and LS to the other Co ion in the unit cell.
%As a result, the MS configuration lowers the crystal symmetry from $R3c$ to $Rc$ by producing the bond-disporportionation. While the crystal volume for MS is similar to the volume of IS state ($V\sim$115.96\AA$^3$),
%the HS site becomes a Co-O long-bond (LB) with $d\sim$1.99\AA and the LS site is a short-bond (SB) with $d\sim$1.94\AA, resulting in the bond-length difference of $\sim$0.05\AA.
The summary of the relaxed structure information is given in Table$\:$\ref{tab:relax}.

\begin{center}
\begin{table}%[!htbp]
\begin{tabular}{ | c | c | c | c | c | c | } 
\hline
Structures & S.G. & Vol.[\AA$^3$] & Co-O[\AA] & Mom.[$\mu_B$] & $N_d$ \\ 
\hline
S1(LS) & $R\bar{3}c$ & 56.40 & 1.95 & 0 & 7.1 \\ 
\hline
S2(IS) & $R\bar{3}c$ & 57.98 & 1.97 & 2.2 & 7.1 \\ 
\hline
S3(MS) Co1 & $R\bar{3}$ & 57.98 & 1.99& 3.0 & 6.9 \\ 
\hline
S3(MS) Co2 & $R\bar{3}$ & 57.98 & 1.94& 0.3 & 7.2 \\ 
\hline
\end{tabular}
\caption{
The structural information of LaCoO$_3$ relaxed structures obtained using DFT+U with U=6eV and J=0.9eV. 
Three different structures are obtained by relaxing with different magnetic moments and
they are denoted as S1 (relaxed with LS), S2 (relaxed with IS), and S3 (relaxed with MS).
The space group (S.G.), the crystal volume per formula unit, the Co-O bond length, magnetic moments, and the $d-$occupancy ($N_d$) are given in this table.
\label{tab:relax}}
\end{table}
\end{center}

\subsection{DFT+DMFT}

Using LS, IS, and MS structures (denoted as S1, S2, and S3 in Table $\:$\ref{tab:relax}) obtained from DFT+U relaxations, we employ a charge-self-consistent DFT+DMFT method~\cite{PhysRevB.90.235103,DMFTwDFT} to study the nature and energetics of the spin-state transition. The DFT+DMFT method is implemented using the maximally localized Wannier functions (MLWFs)~\cite{MOSTOFI20142309,RevModPhys.84.1419} as localized orbitals.
First, we solve the non-spin-polarized Kohn-Sham (KS) equation within DFT using the VASP code.
%Then, we use the maximally localized Wannier functions (MLWFs)~\cite{MOSTOFI20142309,RevModPhys.84.1419} as localized orbitals for the DFT+DMFT method.
Then, we construct Co 3$d$ and O 2$p$ orbitals using MLWFs to represent the hybridization subspace for solving DMFT equations.
The Co-O covalency effect can be treated within DMFT by including both $d$ and $p$ orbitals in the hybridization subspace.
The $p-d$ Hamiltonian of the MLWF basis is constructed from the DFT bands in the hybridization energy window of 11eV. 
Then, the correlated subspace of Co 3$d$ orbitals is treated using the continuous time quantum Monte Carlo (CTQMC)~\cite{PhysRevB.75.155113,RevModPhys.83.349} impurity solver by solving the DMFT self-consistent equations. 
The Hubbard U is 6eV and the Hund's coupling J is 0.9eV within the Co 3$d$ shell while we compute the local self-energy for Co 3$d$ orbitals within DMFT calculations for the study of spin-state transitions. 
Within DMFT, we use temperatures from 100K to 1000K to study the temperature effect on the spectral function while 300K is used for most calculations unless specified otherwise.
%The temperature we used is 300K. 
%while we compute the local self-energy for Co 3$d$ orbitals within DMFT calculations for the study of spin-state transitions.
Here, both U and J are parameterized by Slater integrals ($F^0$, $F^2$, and $F^4$), namely U=$F^0$ and J=$(F^2+F^4)/14$. 
The Coulomb interaction matrix elements with only density-density types are considered in CTQMC while the spin-flip and pair-hopping terms are neglected.
Also, we chose the cartesian axes of Wannier orbitals to be aligned along the Co-O bonds so that the off-diagonal terms in the $d-$Hamiltonian can be much smaller (even zeros) than the on-site terms (see Appendix A).
Therefore, the off-diagonal terms in the DMFT hybridization function can be also ignored within CTQMC.

The charge-self-consistency in DFT+DMFT can be achieved by updating the charge density using the DMFT local Green's function while the DMFT loop is converged. %, the charge density is also recomputed from the DMFT local Green's function and updated. 
The KS equation is solved again using the updated charge density and the new $p-d$ Hamiltonian is constructed using the updated MLWFs obtained from the KS solutions.
%More details of this DFT+DMFT procedure can be found in Ref.$\:$\onlinecite{PhysRevB.90.235103}.
The full charge-self-consistent DFT+DMFT loop is continued until both the charge density, the DMFT Green's function $G^{loc}$, and the DMFT self-energy $\Sigma^{loc}$ are converged.
More details of the DFT+DMFT implementation combining the projected augmented wave method in the DFT part and the formula for the charge update can be found in Ref.~\onlinecite{PhysRevB.90.235103}.
For the precise convergence, the energy cut-off of 600eV and the $k-$point mesh of 8$\times$8$\times$8 are used for DFT loops and
a more dense $k-$point mesh of 30$\times$30$\times$30 is used within the hybridization window of Wannier orbitals for DMFT loops.

Once the DFT+DMFT self-consistency loop is converged, the total energy $E$ is computed using the following formula:
\begin{eqnarray}
E &=& E^{DFT}[\rho]+\frac{1}{N_{\mathbf{k}}}\sum_{\mathbf{k},i\in W}\epsilon_{i\mathbf{k}}\cdot(n_{i\mathbf{k}}-n^0_{i\mathbf{k}}) \nonumber \\
&&+ E^{POT}[G^{loc}]- E^{DC}[N_d] 
\label{eq:E}
\end{eqnarray}
where $E^{DFT}$ is the DFT energy computed using the charge density $\rho$ obtained within DFT+DMFT, $\epsilon_{i\mathbf{k}}$ is the DFT eigenvalues, $n_{i\mathbf{k}}$ is the diagonal part of the DMFT occupancy matrix element with the KS band index $i$ and the momentum $\mathbf{k}$, $W$ is the energy window for the hybridization subspace, and $n^0_{i\mathbf{k}}$ is the DFT occupancy matrix element with the KS band $i$ and the momentum $\mathbf{k}$.
The potential energy $E^{POT}$ within DMFT is given by the Migdal-Galiski formula: $E^{pot}=\frac{1}{2}$Tr$[\Sigma^{loc}(i\omega)\cdot G^{loc}(i\omega)]$.

The double counting (DC) energy, $E^{DC}$ needs to be defined for beyond-DFT methods such as DFT+DMFT and DFT+U since the potential energy treated in the correlated subspace is already accounted as the part of the DFT energy and it needs to be subtracted from the total energy formula. 
Various DFT+DMFT calculations suggest that the DC potential, $V^{DC}(=\partial E^{DC}/\partial N_d)$, smaller than the conventionally used fully-localized-limit (FLL) form~\cite{PhysRevB.44.943} can produce better agreements of energetics~\cite{PhysRevB.89.245133,PhysRevB.90.235103}, the metal-insulator transition~\cite{PhysRevB.90.075136,PhysRevB.86.195136,PhysRevB.89.161113}, and the $p-d$ orbital splitting~\cite{PhysRevB.90.075136,PhysRevB.89.161113,PhysRevLett.115.196403} of oxides compared to experiments. 
Recently, it has been also shown that the more exact form of $V^{DC}$ within DFT+DMFT can be derived~\cite{PhysRevLett.115.196403} and the exact $V^{DC}$ value can be close to the nominal DC form, in which the $d-$occupancy, $N_d$, in the FLL formulae is replaced to the nominal $d-$occupancy, $N_d^0$ ($d^6$ in the LaCoO$_3$ case).
More detailed discussions about different formula of double counting corrections are given in Appendix B.

In this paper, we propose the following form of $V^{DC}$ to allow the change between the FLL form and the nominal DC form by replacing $N_d$ to $\overline{N_d}$:
\begin{eqnarray}
E^{DC}=\frac{U}{2}\cdot\overline{N_d}\cdot(\overline{N_d}-1)-\frac{J}{4}\cdot\overline{N_d}\cdot(\overline{N_d}-2)\\
\label{eq:Edc4}
V^{DC}=U\cdot(\overline{N_d}-\frac{1}{2})-\frac{J}{2}\cdot(\overline{N_d}-1)
\label{eq:Vdc4}
\end{eqnarray}
%where $\overline{N_d}=N_d-\alpha$. 
where U and J are the same paramters which are defined above for the Slater-type interaction, $\overline{N_d}=N_d-\alpha$ where $N_d$ is the $d-$occupancy obtained self-consistently at each correlated site, and $\alpha$ is a parameter which can be tuned for obtaining different $E^{DC}$ and $V^{DC}$ values from the conventional FLL form. 
Our $V^{DC}$ formula can be derived from $E^{DC}$ ($V^{DC}=\partial E^{DC}/\partial N_d$) and allow site-dependent DC potentials.   
Our modified $V^{DC}$ form can recover the conventional FLL DC form by setting $\alpha$=0. 
By increasing $\alpha$, $V^{DC}$ approaches to the nominal $V^{DC}$ value as $\overline{N_d}$ becomes $N_d^0$ ($\alpha=N_d-N_d^0$). 
One should note that changing $V^{DC}$ with different $\alpha$ values can also tune the $p-d$ covalency effect by effectively shifting the $d$ orbital level.
For example, a smaller $V^{DC}$ potential than the FLL DC potential will make the $d$ orbital level higher and the covalency effect weaker, resulting in a reduced $N_d$ value.
The $p-d$ energy separation predicted by DFT or DFT+U with the FLL DC can be usually overbound and the physical role of $\alpha$ is to avoid this overbinding effectively by increasing the $p-d$ energy separation.    
%By increasing $\alpha$, one can approach the nominal DC form when $\alpha=N_d-N_d^0$.
In this paper, We studied the effect of the Co-O covalency on energetics and the nature of spin states in LaCoO$_3$ by changing $V^{DC}$ potentials using different $\alpha$ values.

\section{Results}

\subsection{Energetics of spin-state transition: DFT+DMFT vs DFT+U}
\label{sec:result1}

\begin{figure}[!ht]
\vspace{-0.3cm}
\centering{
\includegraphics[width=0.45\textwidth]{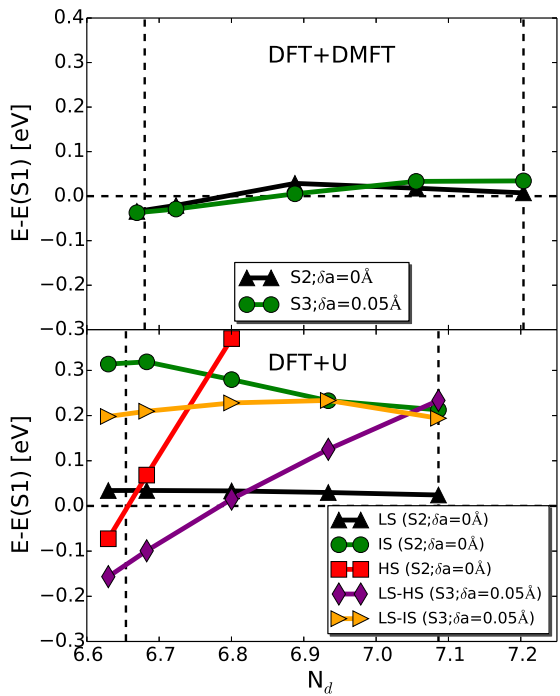}
}
\vspace{-0.5cm}
\caption{Total energy differences between excited spin-state structures (S2 and S3) and the LS structure (S1) in LaCoO$_3$ computed using DFT+DMFT (top panel) and DFT+U (bottom panel) as a function of $N_d$ tuned by different $\alpha$ values in $V^{DC}$ (Eq.$\:$\ref{eq:Vdc4}). 
The S3 structure ($\delta$a=0.05\AA) can incorporate the mixed spin-states between two Co ions in the unit cell.
%The mixed spin-states can incorporate both the volume expansion and the Co-O BD (S3;$\delta$a=0.05\AA$\:$).
%The $y-$axis is the energy difference between the expanded-volume structures with possible spin excitation and the LS volume.
%The N$_d$ values are varied by changing the double counting potential $V^{dc}$ in Eq.$\:$\ref{eq:Vdc} with different $\alpha$ values (0, 0.2, 0.4, 0.6, and 0.7). 
The left vertical dashed line represents the $N_d$ value obtained from the nominal DC formulae while the right dashed line shows $N_d$ obtained using the FLL DC formulae.
%computed using DFT+DMFT (top panel) and DFT+U (bottom panel). The $y-$axis is the $d-$occupancy, N$_d$, and each point is obtained by tuning $\alpha$ of the $V^{dc}$ in Eq.$\:$\ref{eq:Vdc} as 0, 0.2, 0.4, 0.6, and 0.7.
%DFT+U can have multiple solutions for a given $\alpha$ depending on initial spin states.
Both DFT+DMFT and DFT+U methods use the same correlated orbitals (MLWFs) with the same interaction parameters (U=6eV and J=0.9eV). Temperature is 300K within DMFT calculations.
}
\label{fig:fig1}
\end{figure}

Here, we begin by showing the effect of different DC potential ($V^{DC}$) values on the energetics of spin-states in LaCoO$_3$ computed using DFT+DMFT (Fig.$\:$\ref{fig:fig1} top panel) and DFT+U (Fig.$\:$\ref{fig:fig1} bottom panel). % in Fig.$\:$\ref{fig:fig1}.
The $x-$axis shows $N_d$ values obtained by changing $V^{dc}$ in Eq.$\:$\ref{eq:Vdc4} using different $\alpha$ values, namely $\alpha$=0, 0.2, 0.4, 0.6, and 0.7.
Two vertical dashed lines indicate the $N_d$ values obtained using the FLL DC formulae (the right line) and the nominal DC formulae (the left line). The calculations with the FLL DC ($\alpha$=0) converge to $N_d\sim7.2$ for DFT+DMFT and 7.1 for DFT+U while those with the nominal DC result in $N_d\sim6.68$ for DFT+DMFT and 6.65 for DFT+U. 
%The FLL double counting ($\alpha$=0; vertical dashed line) converges to N$_d\sim$7.2 in DFT+DMFT while  N$_d\sim$7.1 is obtained in DFT+U.
The $y-$axis indicates the total energy difference between the expanded volume structures (S2 and S3; $V\sim57.98$\AA$^3$) accompanying excited spin-states and the S1 structure ($V\sim56.4$\AA$^3$).
Here, DFT+U energies are computed by adopting the MLWFs as correlated orbitals consistently with DFT+DMFT calculations, therefore the difference between DFT+DMFT and DFT+U results is attributed purely to the dynamical correlation effect beyond DFT+U.
Also, the previous study of DFT+U calculations using different choices of orbitals (MLWFs vs projectors) has shown that results of the structural phase diagram of nickelates are almost the same as long as the same U and J values are used~\cite{PhysRevB.92.035146}. Therefore we expect that the energetics results reported here will not depend much on the choice of correlated orbitals. 

The energetics obtained using two methods show noticeable differences depending on $N_d$ (the Co-O covalency effect).
The DFT+DMFT energy difference between two expanded volume structures (S2($\delta$a=0\AA)and S3($\delta$a=0.05\AA)) is much smaller (maximally 20meV) than the DFT+U energy difference while DFT+U solutions converge to various meta-stable states for a given structure (LS, IS, and HS for S2 and LS-HS and LS-IS for S3) and the DFT+U energy depends sensitively on $N_d$ and spin-states.
This difference in energetics arises since the spin-state within DFT+DMFT is described as a multi-configurational state with a mixture of various spin states showing smooth crossover (see Fig.$\:$\ref{fig:fig2}) while the DFT+U solution is based on a single-determinant form and produces meta-stable states depending on given structure and $N_d$.
%As a result, DFT+U solutions are converged to various meta-stable states for a given structure and $N_d$ while DFT+DMFT spin-states show a smooth crossover as the structure and $N_d$ change.

Within DFT+DMFT, the S2 structure (triangular dot) is more stable when $N_d>7.0$ and the LS state is dominant while the S3 BD structure (circular dot) becomes stable when the spin-state transition to LS-HS occurs ($N_d<7.0$).
Two structures become energetically almost the same when $N_d$ is further reduced ($N_d\sim6.7$) since the energy of the S2 structure gets lowered as HS is more excited at smaller $N_d$.
The dependence of DFT+U energetics on $N_d$ favors the MS (LS-HS) state although it behaves qualitatively similar to DFT+DMFT as LS with the S2 structure is stable when $N_d>6.8$ and LS-HS with the S3 BD structure becomes rapidly stable when $N_d$ is further reduced.
The energy of HS with the S2 structure is also rapidly decreasing at smaller $N_d$ as the Hund's coupling lowers the energy in the HS state.
Within DFT+U, energetics of other meta-stable states including LS and LS-IS do not depend much on $N_d$ while IS is not favored as $N_d$ is reduced.
Our calculations show that the spin-state transition from the S1 (LS) structure to excited spin-states with the expanded volume $(E-E(S1)<0)$ occurs when $N_d$ becomes smaller ($<6.9$ for DFT+DMFT and $<6.8$ for DFT+U) than the FLL DC result. 
This is consistent with the nominal $V^{DC}$ result (the left vertical dashed lines; $N_d\sim6.68$ for DFT+DMFT) and due to the fact that higher spin states are more excited at smaller $N_d$ as will be shown in Fig.$\:$\ref{fig:fig2}.

\subsection{Nature of paramagnetic states}
\label{sec:result2}

\begin{figure}[!ht]
\vspace{-0.3cm}
\centering{
\includegraphics[width=0.5\textwidth]{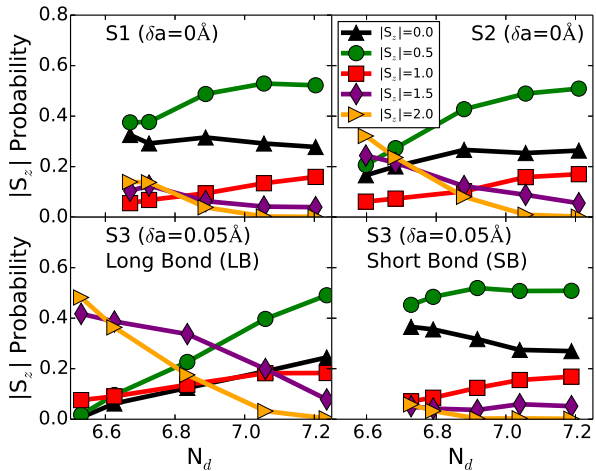}
}
\vspace{-0.5cm}
\caption{The occupation probability of the spin $|S_z|$ state sampled using the CTQMC method within DFT+DMFT as a function of $N_d$ for the S1 structure ($V=56.40$\AA$^3$, upper left panel), the S2 structure ($V=57.98$\AA$^3$, upper right panel), and the S3 structure with the long-bond (lower left panel) and the short-bond (lower right panel) sites.}
\label{fig:fig2}
\end{figure}

Now we turn to the nature of paramagnetic states in LaCoO$_3$ obtained within DFT+DMFT.
Fig.$\:$\ref{fig:fig2} displays the spin $|S_z|$ probabilities sampled using CTQMC as a function of $N_d$ for different structures.
The multi-configurational nature of the paramagnetic state treated in DFT+DMFT means that various $|S_z|$ states ($S_z=0.0\sim2.0$) contribute to the solution.
In all structures, the charge $d^7$ state with $|S_z|=$ 0.5 and 1.5 is not negligible in addition to the nominal $d^6$ state with $|S_z|=$ 0, 1, and 2, therefore LaCoO$_3$ is strongly covalent with the dynamically fluctuating nature of spin and charge states.
%The covalent nature of Co $d$ and O $p$ orbitals means that the charge $d^7$ state with $|S_z|$=0.5 and 1.5 is also strongly populated in addition to the nominal $d^6$ state, therefore both spin and charge states are coupled and strongly fluctuating.
When $N_d>7.0$, LS with $|S_z|=$ 0 and 0.5 has the highest probability for all structures, consistently with the fact that the S1 structure ($V=56.40$\AA$^3$) is energetically stable.
As $N_d$ is reduced, LS is still dominant for the S1 structure while HS with $|S_z|=$ 1.5 and 2 becomes more excited for the S2 structure ($V=57.98$\AA$^3$).
The nature of paramagnetic state in the S2 structure, which is energetically stable when $N_d<6.7$, is characterized by a mixture of both LS and HS, fluctuating dynamically with non-negligible probabilities. 
The IS state ($|S_z|=$1.0) is strongly suppressed for all structures.
In the BD structure, HS with $|S_z|=$ 1.5 and 2 becomes rapidly favored in the LB site as $N_d$ is reduced while LS with $|S_z|=$ 0 and 0.5 is always dominant in the SB site.
These DFT+DMFT results of paramagnetic states exhibiting the mixture of different spin states are distinct from the DFT+U results where meta-stable solutions of spin states are found.

\subsection{Charge ordering induced by spin-state ordering}
\label{sec:result3}

\begin{figure}[!ht]
\vspace{-0.3cm}
\centering{
\includegraphics[width=0.45\textwidth]{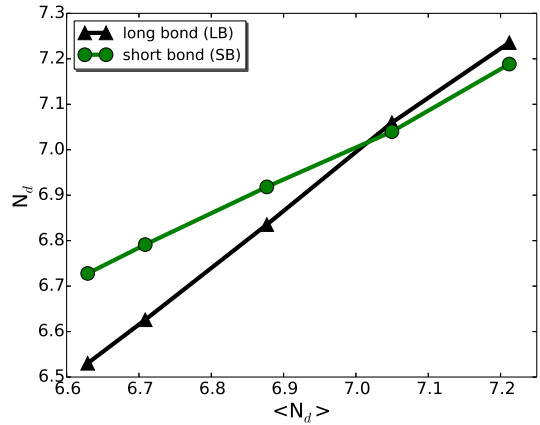}
}
%\vspace{-0.5cm}
\caption{The $d-$occupancy, $N_d$ computed for both the long-bond and the short-bond sites in the S3 structure computed using DFT+DMFT as a function of the average $\langle N_d \rangle$ between two sites. %The y-axis is $N_d$ for both LB and SB sites. The x-axis is the averaged $N_d$. C.S.C and N.C.S.C. results are compared}
}
\label{fig:fig3}
\end{figure}

%\textit{Charge-ordering:}
The strongly coupled spin and charge degrees of freedom also produce an intriguing rock-salt type charge ordering state induced from the MS state in the S3 structure. 
%Our DFT+DMFT calculation shows that the spin-state ordering occurring in the BD structure can also induce the charge ordering.
Fig.$\:$\ref{fig:fig3} shows that the DFT+DMFT charge in the LB site (triangular dots) gets smaller more rapidly than the SB charge (circular dots) as the average $\langle N_d \rangle$ becomes reduced (the overall Co-O covalency is reduced).
This is because HS in the LB site favors the $|S_z|=2$ with $d^6$ state while the $|S_z|=0.5$ with $d^7$ state is dominant in the SB site as the Co-O covalency remains strong.
This enhancement of charge ordering when $N_d<$6.9 is also consistent with the rapid increase of the HS probability in the LB site as shown in Fig.$\:$\ref{fig:fig2}.
The experimental evidence of charge ordering with the concomitant spin-state ordering in LaCoO$_3$ has been reported in the tensile-strained film~\cite{PhysRevLett.120.197201}.
The ground-state of this charge ordered LaCoO$_3$ is insulating and the nature of the LB state is a Mott insulator while the SB state is a band insulator due to the strong Co-O hybridization (see Fig.$\:$\ref{fig:fig8}). 
This MS state in the S3 (BD) structure is somewhat reminiscent of the site-selective Mott physics occurring in nickelates in which the LB site is Mott insulating with the $d^8$ state while the SB site is a covalent insulator hybridized with O hole states as $d^8L^2$.~\cite{PhysRevLett.109.156402}
Although charge ordering between two Ni sites in nickelates is not important to induce the insulating state in nickelates, charge ordering ($\sim0.2$) in LaCoO$_3$ is naturally induced from the spin-state ordering.
Also, nickelates are negative charge-transfer insulators meaning that almost one electron is donated to each Ni ion from surrounding O ions
while the Co ion in LaCoO$_3$ favors a mixed-valence state with $N_d=6.6\sim6.7$ as the charge transfer from the O hole is smaller than nickelates.
%Nickelates are negative charge-transfer insulators, and exactly one electron is donated to each Ni ion from surrounding O ions.
%The site-selective Mott insulating phase occurs as the LB site is Mott insulating with the $d^8$ state while the SB site is a covalent insulator with the $d^8L^2$ state. 
%As a result, charge ordering between two Ni sites is not necessary to induce the insulating state.
%However, charge ordering ($\sim0.2$) should be accompanied by the mixed spin-state with the Co-O breathing distortion in LaCoO$_3$.
%And the energetics of the spin-state transition are consistent when the Co ion is a mixed valent state ($N_d=6.6\sim6.7$). 
%meaning that LaCoO$_3$ is away from the negative charge-transfer regime than nickelates.

%However, LaCoO$_3$ is not as close to the negative charge-transfer regime as nickelates, and the spin-state transition occurs when $N_d=6.6\sim6.7$.
%Also charge ordering ($\sim0.2$) is always induced by the mixed spin-state with the Co-O breathing distortion.

%The charge ordering is pronounced as $\langle N_d \rangle$ becomes smaller than 6.9 where the HS-LS spin-state ordering also occurs in Fig.$\:$\ref{fig:fig2}.
%The reduction of the charge in the LB site is consistent with the increased $|S_z|=2.0$ state ($d^6$) dominant in the HS state while the SB site favors the $|S_z|=0.5$ with a $d^7$ configuration as the Co ion is still covalently bonded with the O ion.

\subsection{Density of states}
\label{sec:result4}

\begin{figure}[!ht]
\vspace{-0.3cm}
\centering{
\includegraphics[width=0.45\textwidth]{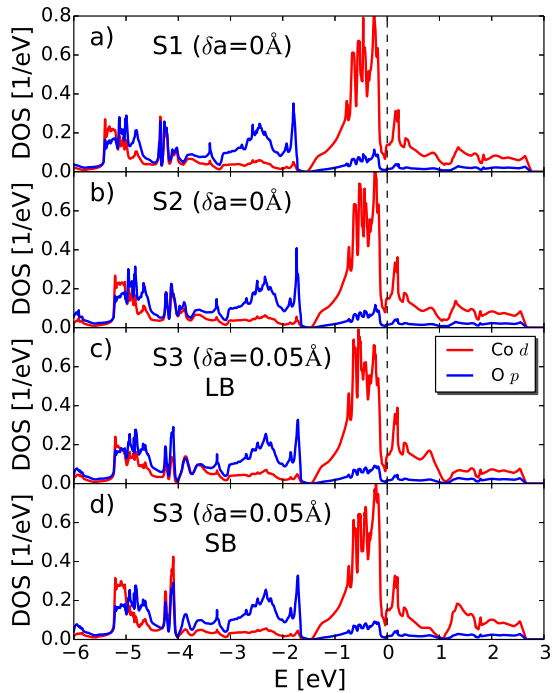}
}
\vspace{-0.5cm}
\caption{The density of states computed for different structures of LaCoO$_3$ using DFT with the paramagnetic (no spin-polarization) symmetry. They are a) S1 ($V=$56.40\AA$^3$), b) S2 ($V=$57.98\AA$^3$), and S3 ($V=$57.98\AA$^3$) structures with c) the Co-O long bond (LB) site and d) the short bond (SB) site. 
%The $N_d$ value is 6.67 obtained using $\alpha=0.7$ for $V^{DC}$
}
%In DFT+DMFT, the Hubbard interaction U is 6eV, the Hund's coupling J is 0.9eV, temperature is 300K, and the double counting parameter $\alpha$ is 0.7.}
%the $R3c$ structure relaxed using DFT+U IS state compared for $N_d\sim 7.05$ (top panel) and $N_d\sim 6.7$ (bottom panel)}. (b) The DOS computed using DFT+DMFT for the BD structure.
\label{fig:fig4}
\end{figure}

In this subsection, we study the correlation, structure, $p-d$ covalency, and temperature effects on the spectral functions of LaCoO$_3$.
First, we show the density of states (DOS) computed using DFT paramagnetic calculations (no spin polarization) in Fig.$\:$\ref{fig:fig4}.
The DOS computed for the different structures shows very similar features. 
All structures are metallic without correlations and the S1 structure has slightly larger band-width than other structures due to the smaller crystal volume. 
Co $t_{2g}$ states are all occupied below the Fermi energy Co $e_g$ states are widely spread due to the strong mixing with the O $p$ states. 
The LB and SB sites in the S3 structures exhibit the similar DOS without charge ordering ($N_d\sim 7.2$).

\begin{figure}[!ht]
\vspace{-0.3cm}
\centering{
\includegraphics[width=0.45\textwidth]{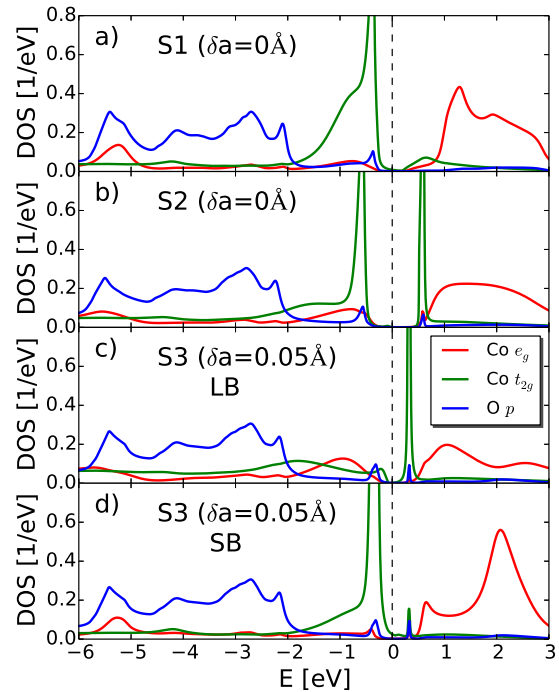}
}
\vspace{-0.5cm}
\caption{The density of states obtained for different structures of LaCoO$_3$ using DFT+DMFT with U=6eV, J=0.9eV, and the DC parameter $\alpha=0.7$ (Eq.$\:$\ref{eq:Vdc4}) resulting $N_d\sim6.68$. Different structures are a) S1 ($V=$56.40\AA$^3$), b) S2 ($V=$57.98\AA$^3$), and S3 with c) the Co-O long bond and d) the short bond sites.
%Different structures are the (a) LS volume (56.40\AA$^3$), (b) the expanded volume (57.98\AA$^3$), and both (c) LB and (d) SB sites for the BD structure.
%The $N_d$ value is 6.67 obtained using $\alpha=0.7$ for $V^{DC}$
}
%In DFT+DMFT, the Hubbard interaction U is 6eV, the Hund's coupling J is 0.9eV, temperature is 300K, and the double counting parameter $\alpha$ is 0.7.}
%the $R3c$ structure relaxed using DFT+U IS state compared for $N_d\sim 7.05$ (top panel) and $N_d\sim 6.7$ (bottom panel)}. (b) The DOS computed using DFT+DMFT for the BD structure.
\label{fig:fig5}
\end{figure}

Now, we include strong correlation effects in three structures (S1, S2, and S3) within DFT+DMFT and compute the DOS using the paramagnetic spin configuration in Fig.$\:$\ref{fig:fig5}. 
We see dramatic changes of electronic structures due to correlation effects as well as the structural changes from the volume expansion and the Co-O bond disproportionation. 
%Fig.$\:$\ref{fig:fig5} shows the DFT+DMFT DOS computed using the paramagnetic configuration.  
The DOS in the S1 structure (Fig.$\:$\ref{fig:fig5} a) shows that the band gap is almost 0.6eV, which is consistent with the optical gap measurement~\cite{PhysRevB.46.9976}. The $t_{2g}$ state is almost occupied while the $e_g$ state is mostly unoccupied, as expected for the LS state.
As the volume is expanded, the spin-state transition to higher spins occurs continuously and the $t_{2g}$ state begins to be unoccupied while more $e_g$ orbitals are occupied (Fig.$\:$\ref{fig:fig5} b). 
%the feature of DOS changes such that the $t_{2g}$ orbital is more unoccupied while the $e_g$ state is more occupied demonstrating the spin-state transition from LS to IS-like state.
%Both $e_g$ and $t_{2g}$ orbitals are gapped and the gap size has increased from the LS gap in the S2 structure.
%The nature of the paramagnetic insulating state is a Mott insulator, which originates from the emergence of the pole in the imaginary part of the self-energy due to the increased HS state (see the next subsection).
%The charge gap size has also increased compared to the LS gap.
In the S1 structure, the major optical transition occurs from the Co $t_{2g}$ valence band peak to the Co $e_g$ conduction band peak while the transition between the same $t_{2g}$ bands dominates in the S2 structure. Therefore, the position of the first major peak in the optical conductivity will be reduced as the crystal volume expands at higher temperature, which is consistent with the optical conductivity measurement in experiment. Moreover, our calculation shows that the S3 structure with the MS state produces the smaller gap than the S2 structure, therefore the optical gap can be further reduced as temperature is raised presumably with more populated MS states.
The nature of the paramagnetic insulating state in the S2 structure is a strongly correlated band insulator driven by both the Co-O hybridization due to the covalency and the electron localization of the Co d orbitals due to increased HS states. This physics is captured in the imaginary part of the self-energy (see Fig.$\:$\ref{fig:fig8} b) showing the emergence of the sharp pole outside the hybridization gap near the Fermi energy. This paramagnetic insulating state is represented as a mixture of fluctuating HS and LS states (see Fig.$\:$\ref{fig:fig2}) and it is distinct from the normal Mott insulator driven by the pure electron localization which is expected from the HS state in a d6 configuration.
%Also, this paramagnetic insulating state is represented as a mixture of fluctuating HS and LS states due to the Co-O covalency and is distinct from the Mott insulator with a pure HS state expected in a $d^6$ configuration. 
%which has been discussed in literatures as the consquence of the large hybridization between Co and O.
%Although DOS is similar to a IS state, the spin-state is a mixture of dynamically fluctuating HS and LS states while the $|S_z|=1$ state (typically thought as the IS state) is not favored.
In the S3 structure, the LB site becomes higher spin state with more unoccupied $t_{2g}$ states while the SB site remains as LS with a similar gap size as the homogeneous LS gap.  
Here, the $\alpha$ value in $V^{DC}$ (Eq.$\:$\ref{eq:Vdc4}) is set to 0.7 resulting $N_d\sim6.68$ for all structures.
Different $V^{DC}$ values will change the relative position of the O $p$ peak from the Fermi energy as $N_d$ is also varied.
Our O $p$ top peak position is located at -2eV when $N_d\sim6.68$ and this peak position is consistent with the measured X-ray photo-emission spectra~\cite{PhysRevB.47.16124} validating the $\alpha$ value we used.  
%The gap size becomes similar to the LS gap in a LS volume.

\begin{figure}[!ht]
\vspace{-0.3cm}
\centering{
\includegraphics[width=0.45\textwidth]{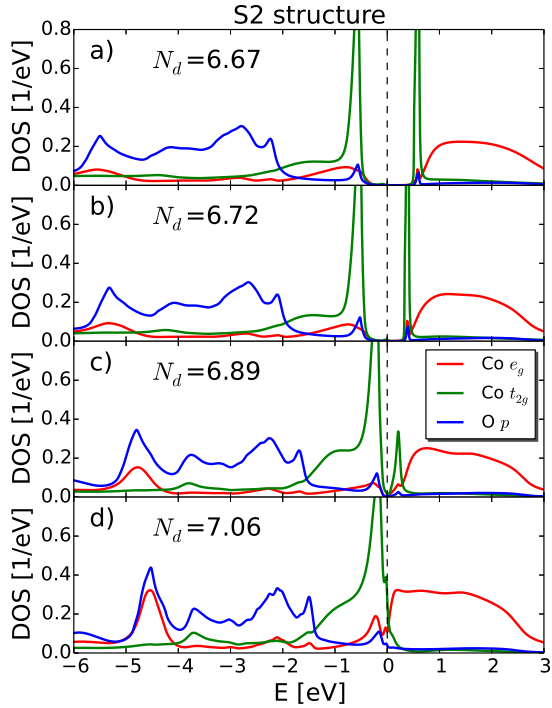}
}
\vspace{-0.5cm}
\caption{The density of states computed using different $\alpha$ values in Eq.$\:$\ref{eq:Vdc4} within DFT+DMFT. The S2 structure is used.
Different $\alpha$ values lead to distinct $N_d$ results, namely a) $N_d=6.67$ ($\alpha=0.7$), b) $N_d=6.72$ ($\alpha=0.6$), c) $N_d=6.89$ ($\alpha=0.4$), and $N_d=7.06$ ($\alpha=0.2$).
U=6eV and J=0.9eV are used within DMFT.
%($\alpha=0.7$ in Eq.$\:$\ref{eq:Vdc4}). Different structures are the (a) LS volume (56.40\AA$^3$), (b) the expanded volume (57.98\AA$^3$), and both (c) LB and (d) SB sites for the BD structure.
}
\label{fig:fig6}
\end{figure}

To clarify the role of the Co-O covalency (parametrized by $N_d$) on electronic structure, we show the DOS obtained using different $\alpha$ values for the S2 structure in Fig.$\:$\ref{fig:fig6}.
As the Co-O covalency is enhanced ($N_d$ is increased), the O $p$ peaks move closer to Co $d$ states near the Fermi energy.
As a result, the spectral gap becomes smaller due to the larger Co-O hybridization and the ground-state is eventually metallic when $N_d\sim 7.1$.
Therefore, the insulator-to-metal transition occurs as the correlation in the Co ion is reduced due to the increased Co-O covalency and, at the same time, the spin states are less excited as Co $t_{2g}$ states are more occupied and $e_g$ states are more unoccupied. 
This DOS result is also consistent with the occupation probability data of the S2 structure (Fig.$\:$\ref{fig:fig2} upper right panel) showing that the HS states ($|S_z|=1.5$ and 2.0) are suppressed and the LS and IS states ($|S_z|=0.0$, 0.5, and 1.0) are gradually increasing as $N_d$ increases.
Our results suggest that the Co-O hybridization due to the covalency can play an important role in explaining the metal-insulator transition and the spin-state transition in LaCoO$_3$.
%As shown in the occupation probability (Fig.$\:$\ref{fig:fig2}), the S2 structure exhibits the spin-state transition as a function of $N_d$. 
%As $N_d$ increases, the HS states ($|S_z|=1.5$ and 2.0) are suppressed and the LS and IS states ($|S_z|=0.0$, 0.5, and 1.0) are gradually increasing.
%The enhanced Co-O covalency  

\begin{figure}[!ht]
\vspace{-0.3cm}
\centering{
\includegraphics[width=0.45\textwidth]{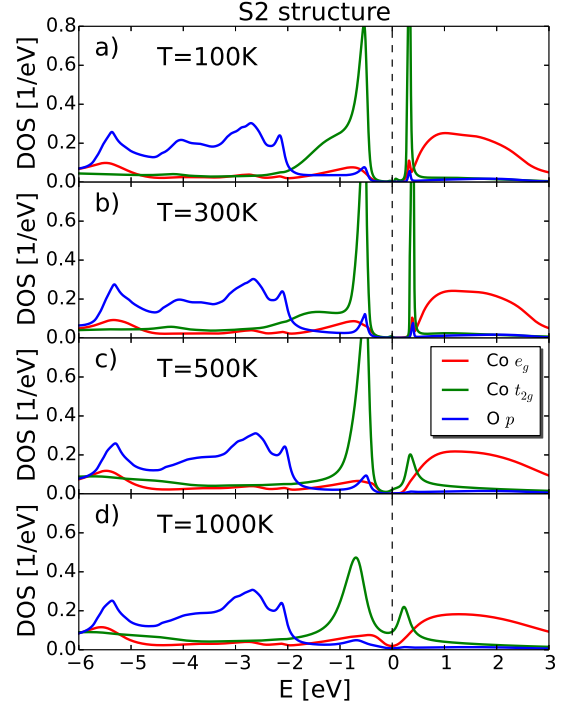}
}
\vspace{-0.5cm}
\caption{The density of states computed for the S2 structure of LaCoO$_3$ using DFT+DMFT at different temperatures, namely a) $T=100K$, b) $300K$, c) $500K$, and d) $1000K$. U=6eV, J=0.9eV, and the double counting $\alpha=0.6$ ($N_d\sim6.7$) are used within DMFT. 
}
\label{fig:fig7}
\end{figure}

Until now, our DFT+DMFT calculations have been performed at the fixed temperature $(\sim300K)$.
Experimentally, LaCoO$_3$ also exhibits the insulator-to-metal transition as temperature is raised above near $T=400K$ but the nature of this transition has not been clarified.
To better understand the role of temperature on the metal-insulator transtion in LaCoO$_3$, we plot the DOS of LaCoO$_3$ at the fixed S2 structure and different temperatures (Fig.$\:$\ref{fig:fig7}).
The DOS data computed at both $100K$ and $300K$ show similar features although the spectral gap at $300K$ is slightly larger than one at $100K$.
The similarity of the DOS at low temperatures indicates that our spin-state calculations at $300K$ can be similarly reproduced at $100K$ (the experimental $T_c$ of the spin-state transition).
More importantly, the strong variation of the DOS and spin-states depending on structures in Fig.$\:$\ref{fig:fig5} means that the structural changes should be incorporated for the better description of electronic structure while temperatures are varied.
Nevertheless, the spectral gap becomes smaller as temperature is raised above $500K$ and the Co $t_{2g}$ state becomes the incoherent metallic state at around $1000K$. 
%Therefore, the critical temperature of this insulator-to-metal transition within our DFT+DMFT calculation is overestimated than the experimental one possibly due to the mean-field nature of the single-site DMFT approximation.     

\subsection{Self-energy data}
\label{sec:result5}

\begin{figure}[!ht]
\centering{
\includegraphics[width=0.45\textwidth]{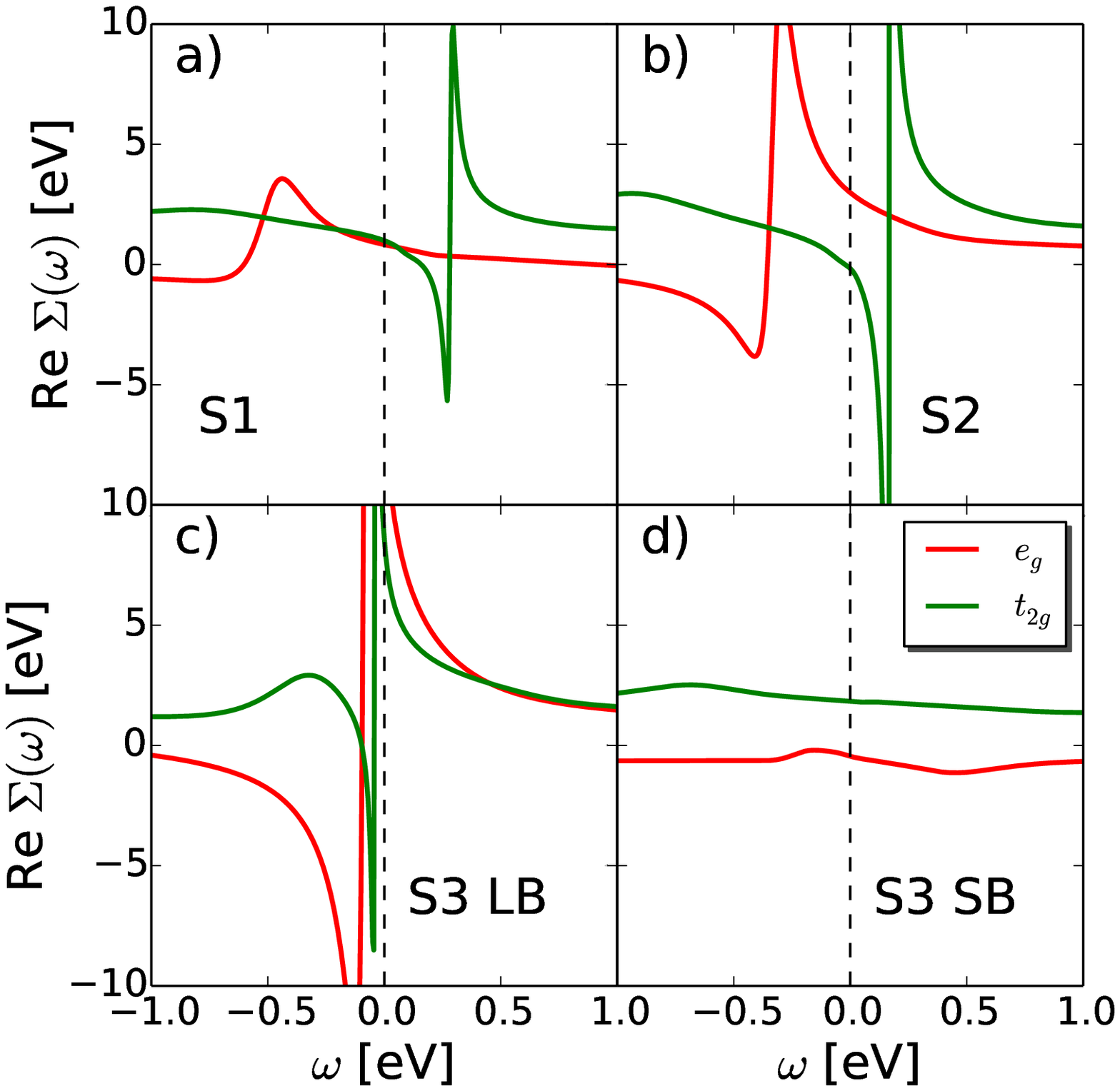}
\includegraphics[width=0.45\textwidth]{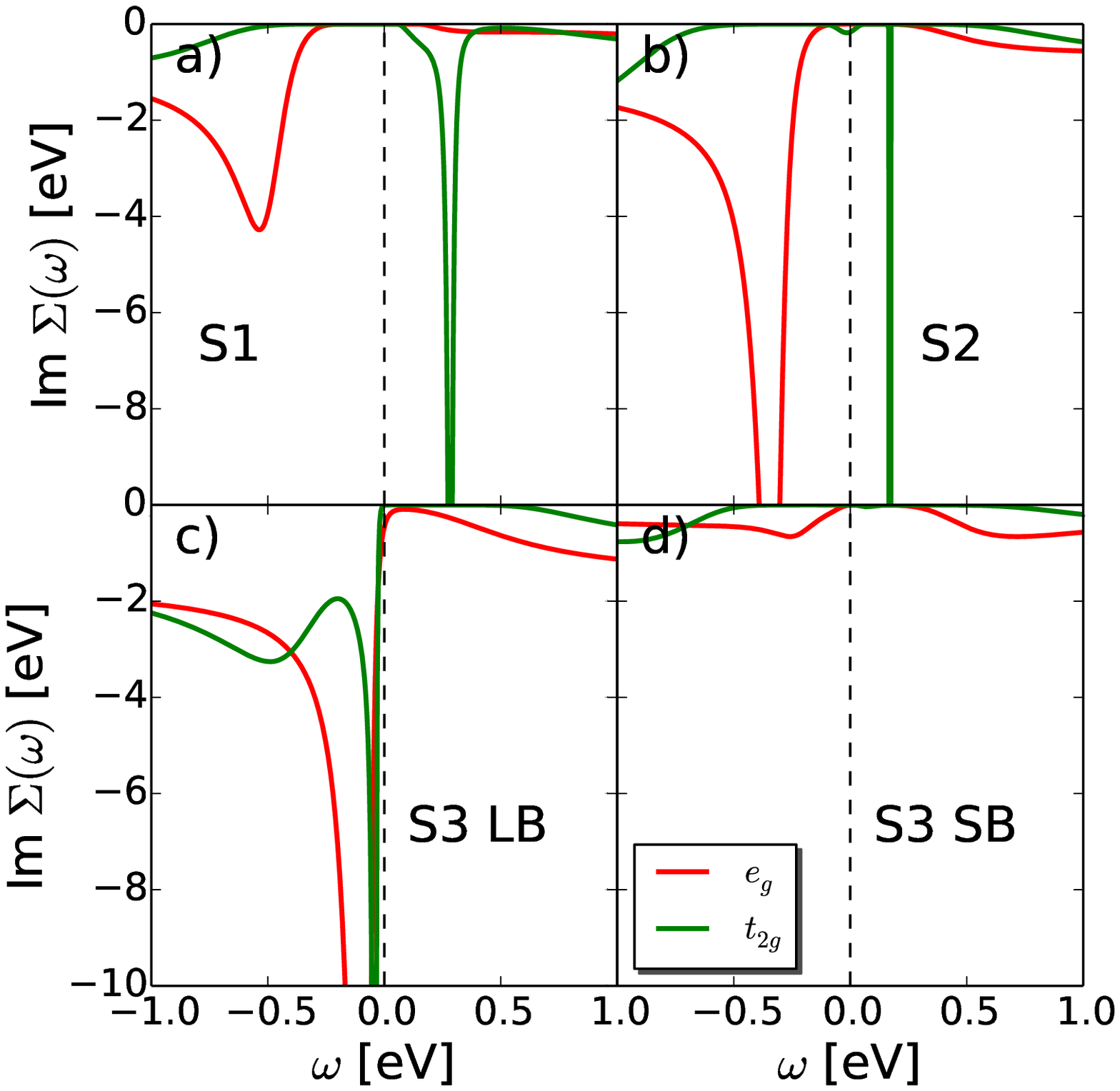}
}
\caption{
The real part (top panel) and the imaginary part (bottom panel) of self-energy data computed for different structures in LaCoO$_3$, namely a) S1, b) S2, c) S3 Co-O long bond (LB), and d) S3 Co-O short bond (SB).
U=6eV, J=0.9eV, and the double counting parameter $\alpha=0.6$ ($N_d\sim6.7$) are used. 
}
\label{fig:fig8}
\end{figure}

To study the nature of insulating states occuring in different stuctures, we show both the real part ($Re\Sigma$; Fig.$\:$\ref{fig:fig8} top panel) and the imaginary part ($Im\Sigma$; Fig.$\:$\ref{fig:fig8} bottom panel) of self-energies on the real axis, which are used to compute the DOS in Fig.$\:$\ref{fig:fig5}.
The S1 structure (Fig.$\:$\ref{fig:fig8}a) exhibits rather small $Im\Sigma$ values in both $e_g$ and $t_{2g}$ orbitals when $\omega\sim 0$ while a sharp pole develops at $\omega\sim 300meV$ in the $t_{2g}$ orbital for both real and imaginary parts. 
This diverging nature of the self-energy indicates Co $d$ orbitals are still correlated even at the S1 structure with LS. 
%This self-energy nature of LS is different from the typical band insulating nature of LS which is expected in a typical $d^6$ configuration. 
This self-energy nature of LS is different from the typical band insulating nature of LS which is expected in a $d^6$ configuration. The nature of our insulating state is the correlated band insulator driven by both the Co-O hybridization and the electron correlation. The electron correlation is encoded in the sharp and narrow pole structure of the imaginary part of the self-energy developed outside the hybridization gap while the imaginary part of the self-energy is still zero at the Fermi energy.
As the structure changes from S1 to S2 along with the volume expansion, the strength of poles becomes enhanced and the positions of the poles get close to the Fermi energy in both $e_g$ and $t_{2g}$ orbitals. This clearly shows that correlations are enhanced due to the increased higher spin probabilities (see Fig.$\:$\ref{fig:fig2}) as the volume is expanded in LaCoO$_3$. 
The insulating nature of the S3 structure shows the ``site-selective'' Mott physics, as the Co-O LB site undergoes a Mott transition with the diverging self-energies at the Fermi energy while the SB site behaves as a band insulator with the small imaginary part and the flat real part of self-energies originated from the strong Co-O covalency.

\begin{figure}[!ht]
\centering{
\includegraphics[width=0.45\textwidth]{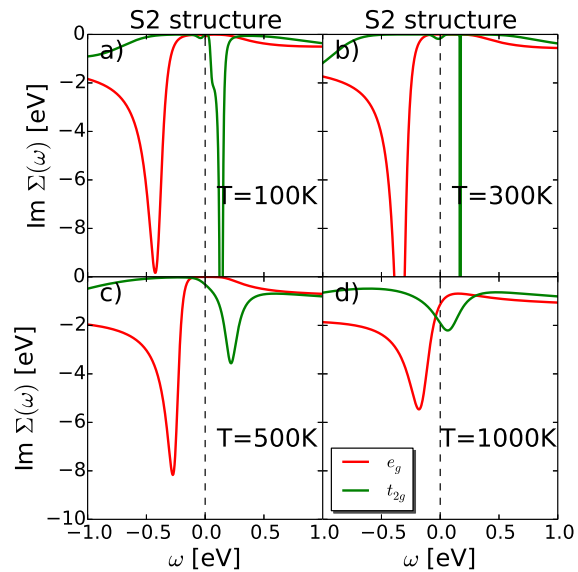}
}
\caption{
The imaginary part of self-energy data computed for different temperatures at a fixed S2 structure in LaCoO$_3$, namely a) $T$=100K, b) $T$=300K, c) $T$=500K, and d) $T$=1000K.
U=6eV, J=0.9eV, and the double counting parameter $\alpha=0.6$ ($N_d\sim6.7$) are used. 
}
\label{fig:fig9}
\end{figure}

To understand the nature of the insulator-to-metal transition in LaCoO$_3$ as temperature is raised, we also plot the $Im\Sigma$ computed for the S2 structure as a function of temperature in Fig.$\:$\ref{fig:fig9}. 
Results of $Im\Sigma$ at $T$=100K show that poles are developed due to correlations at $\omega=$ -450meV for the $e_g$ orbital and at $\omega=$ 150meV for the $t_{2g}$ orbital. 
%The strengths of these poles are enhanced at $T$=300K as the S2 structure can occupy higher spin states at this temperature. 
The strengths of these poles are enhanced at $T$=300K as the S2 structure can occupy higher spin states due to the spin-state transition and electron localization can be enhanced at this temperature. This physics is different from the typical Mott insulator without any spin-state transition, in which the electron localization is usually stronger at lower temperature.
As the temperature is raised even above 300K, the pole strengths are reduced and correlations become weaker for both $e_g$ and $t_{2g}$ orbitals resulting the metallic phase obtained in Fig.$\:$\ref{fig:fig7}.

\subsection{Charge-self-consistency effect}
\label{sec:result6}

\begin{figure}[!ht]
\centering{
\includegraphics[width=0.45\textwidth]{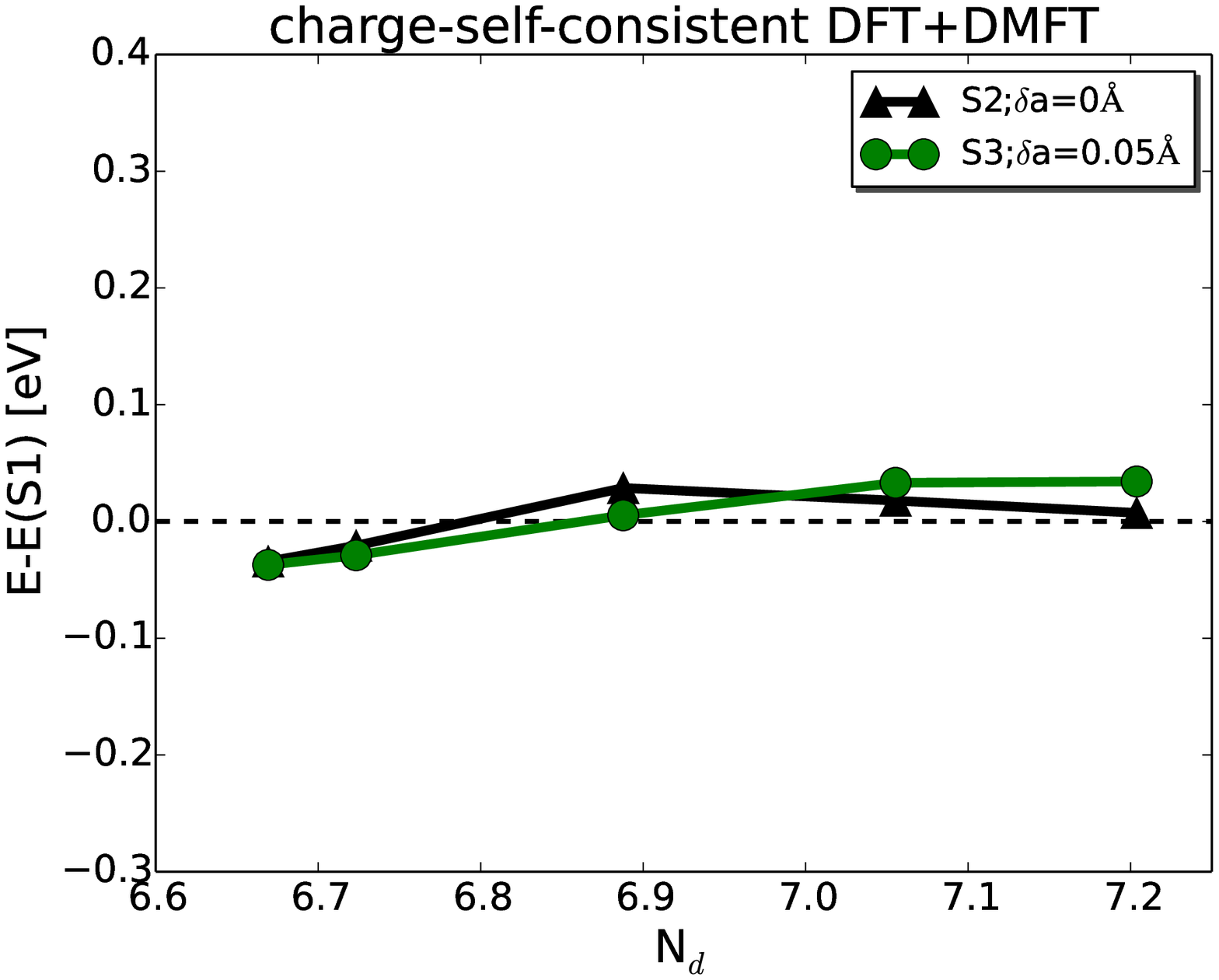}
\includegraphics[width=0.45\textwidth]{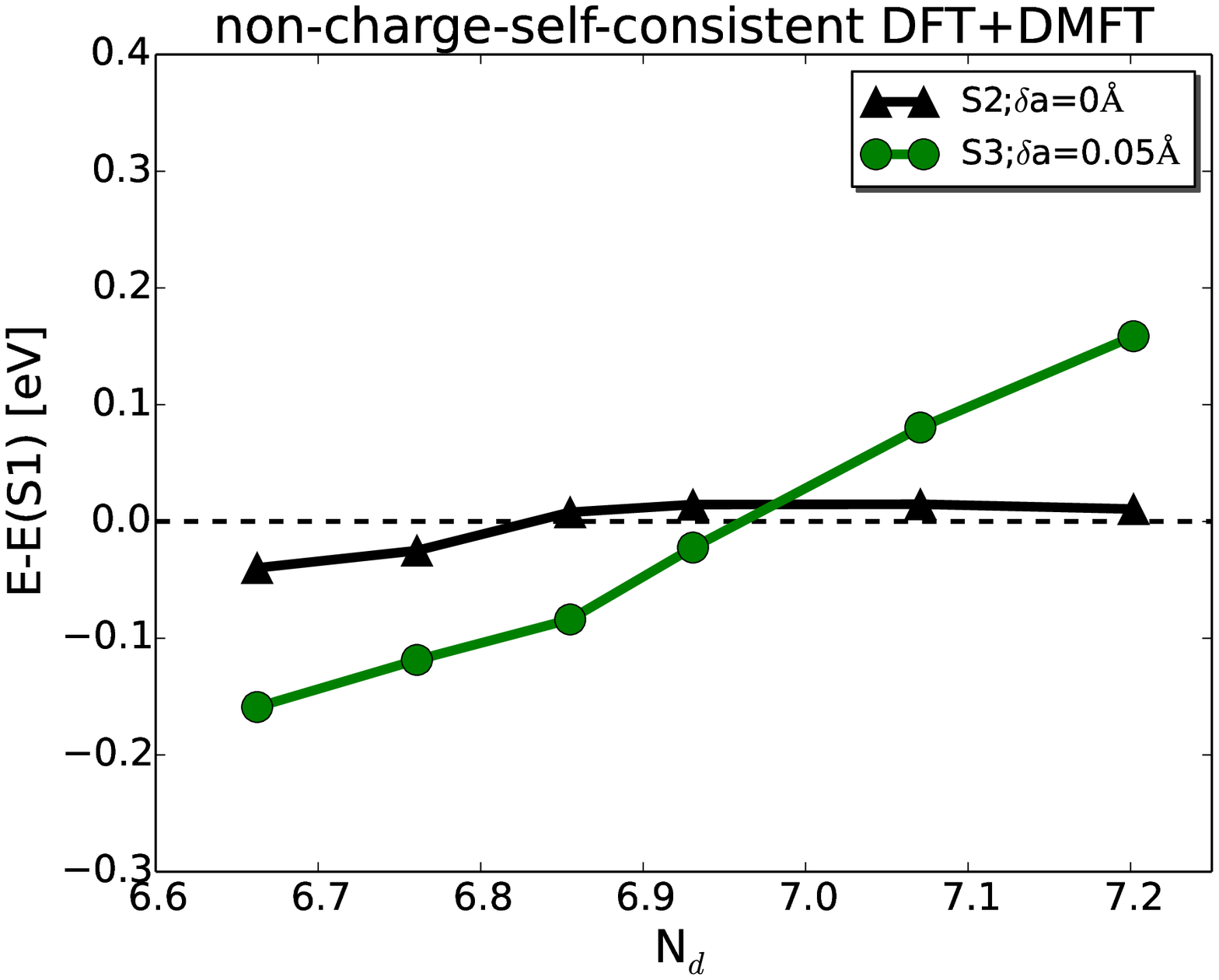}
}
\caption{
Total energy differences between excited spin states (S2 and S3) and the low spin state (S1) in LaCoO$_3$ computed using both charge-self-consistent DFT+DMFT (top panel) and non-charge-self-consistent DFT+DMFT (bottom panel) as a function of $N_d$ tuned by different $V^{DC}$ potentials. The excited spin states incorporates the volume expansion with or without the Co-O bond-disproportionation ($\delta$a=0.05\AA$\:$or 0\AA).
U=6eV, J=0.9eV, and temperature T=300K are used within DMFT.
}
\label{fig:fig10}
\end{figure}

Finally, we show the charge-self-consistency effect in DFT+DMFT on the energetics and electronic structure in LaCoO$_3$. Fig.$\:$\ref{fig:fig10} shows the energetics of spin-state transition in LaCoO$_3$ as a function of $N_d$ comparing charge-self-consistent (top panel) and non-charge-self-consistent (bottom panel) DFT+DMFT calculations.
Here, the non-charge-self-consistent calculation means that the charge density ($\rho$ in Eq.$\:$\ref{eq:E}) is fixed to the DFT one while the DMFT local Green's function ($G^{loc}$ in Eq.$\:$\ref{eq:E}) is obtained by converging DMFT self-consistent equations. 
As a result, the $V^{DC}$ potential is fixed during the DFT+DMFT loop since the charge-density is not updated and $V^{DC}$ is a function of the charge-density ($N_d$). 
%during the loop and $V^{DC}$ is a function of the charge-density ($N_d$).
Therefore, different $N_d$ results in non-charge-self-consistent DFT+DMFT (bottom panel) are obtained by shifting the $V^{DC}$ potential as data points are changed. 
However, in charge-self-consistent DFT+DMFT (top panel), $V^{DC}$ is computed using the Eq.$\:$\ref{eq:Vdc4} with the self-consistently determined charge density ($N_d$) and the corresponding $N_d$ is obtained from the DMFT Green's function.
%A fixed $V^{DC}$ value is used for each data point within the non-charge-self-consistent DFT+DMFT (bottom panel) since the charge-density is not updated and the constant shift of the $V^{DC}$ potential converges to different $N_d$ results while the DC potential $V^{DC}$ is computed using the Eq.$\:$\ref{eq:Vdc4} with the self-consistently determined $N_d$ in the charge-self-consistent DFT+DMFT (top panel). 
%
%The DC potential $V^{DC}$ is computed using the Eq.$\:$\ref{eq:Vdc4} with the self-consistently determined $N_d$ in the C.S.C. DFT+DMFT (top panel) while the fixed $V^{DC}$ value ($V^{DC}$ is fixed during DFT+DMFT since the charge-density is not updated) is used for each data point and the constant shift of the $V^{DC}$ potential converges to different $N_d$ results within the N.C.S.C. DFT+DMFT (bottom panel).
%Different $\alpha$ values used in the Eq.7 generate different $N_d$ results in the left panel while the constant shift of the $V^{DC}$ potential converges to different $N_d$ results in the right panel.    

The energetics of the S2 structure (homogeneous spin states) show the very similar behavior as a function of $N_d$ between charge-self-consistent and non-charge-self-consistent calculations.
In the case of the MS state for the S3 structure, non-charge-self-consistent DFT+DMFT energetics overestimate the tendency toward the spin-state ordering as the energy difference between the excited spin-state and the low-spin (S1) state becomes much lower ($\sim$-150meV) than the charge-self-consistent energetics ($\sim$-30meV) when $N_d\sim$6.7.
The energetics of the MS state also depend on $N_d$ much sensitively in the non-charge-self-consistent case.
Therefore, the main effect of charge-self-consistency in DFT+DMFT is to reduce the spin-state ordering effect in the S3 structure and the energetics between homogeneous and MS spin-states in the S2 and S3 structures become very close.
%Therefore, the effect of the charge-self-consistency within DFT+DMFT can be important for obtaining the energetics of inhomogeneous electronic phases in which spin-state or charge orderings can occur.

\begin{figure}[!ht]
\centering{
\includegraphics[width=0.45\textwidth]{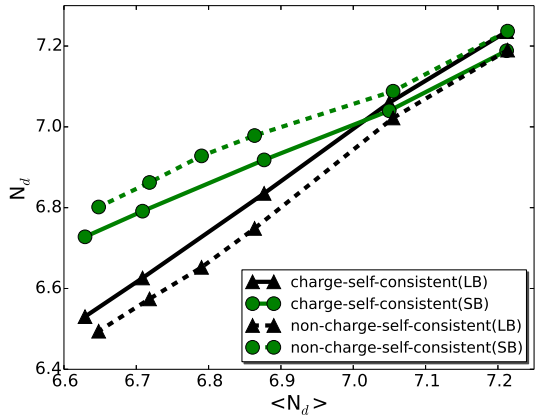}
}
\caption{
The $d-$occupancy, $N_d$ computed for both LB and SB sites in the S3 structure computed using charge-self-consistent DFT+DMFT (solid lines) and non-charge-self-consistent DFT+DMFT (dashed lines) as a function of the average $\langle N_d \rangle$ between two sites. %The y-axis is $N_d$ for both LB and SB sites. The x-axis is the averaged $N_d$. C.S.C and N.C.S.C. results are compared}
}
\label{fig:fig11}
\end{figure}

To further investigate the effect of charge-self-consistency on electronic structure of LaCoO$_3$, we display the $d-$occupancy, $N_d$ in Fig.$\:$\ref{fig:fig11} obtained for both LB and SB sites in the S3 structure computed using charge-self-consistent DFT+DMFT (solid lines) and non-charge-self-consistent DFT+DMFT (dashed lines) as a function of the average $N_d$ between two sites. 
Without the charge update (dashed lines), the $N_d$ difference between two sites is more enhanced since the SB site occupies more $d-$orbitals while the LB site takes even less $N_d$ compared to the charge-self-consistent results (solid lines) across different $\langle N_d \rangle$ values.
Therefore, charge-self-consistency within DFT+DMFT reduces the tendency toward both spin-state and charge orderings between correlated Co sites, as a result, the energy difference between different excited spin-states is also much decreased within the charge-self-consistent calculation.

Here, we fix the U and J values during the DFT+DMFT calculations, therefore the main difference between charge-self-consistent and non-charge-self-consistent results is originated from the change of the Wannier Hamiltonian due to the charge update (see Appendix A). 
The main role of the charge update is to decrease the crystal field splitting in the S1 and S2 structures and to promote the homogeneous spin-state transition. 
At the same time, the spin-state and charge ordering effects in the S3 structure has been reduced as the crystal field splitting in the LB site has been increased while it has been decreased in the SB site acting counterintuitively on the structural effect. 
Moreover, the overall $d-$orbital level in the LB site is substantially lowered than the level in the SB site due to the charge update, again decreasing the charge ordering effect.
Therefore, the charge-self-consistent effect compensates for the spin-state and charge ordering effects in the S3 structure while it favors the homogeneous spin-state transition, and as a result, the energy difference between the spin-state ordering and the homogeneous spin state has been much reduced within the charge-self-consistent DFT+DMFT calculations. 

%\subsection{The effect of different correlated orbitals}
%\label{sec:result5}

\section{Conclusion}
\label{sec:conclusion}

In conclusion,
we adopt the charge-self-consistent DFT+DMFT method to study the nature and energetics of both homogeneous and MS states in covalent LaCoO$_3$.
% incorporating possible lattice responses during the spin-state transition.
We find that structural changes during the spin-state transition are important to understand both energetics and electronic structure in LaCoO$_3$. 
%As the crystal volume is expanded, the occupation probability of higher spin state in a Co ion increases and the ground-state becomes a Mott insulating state as the self-energy diverges. 
As the crystal volume is expanded, the occupation probability of higher spin state in a Co ion increases and a sharp and narrow peak in the imaginary part of the self-energy develops outside the Co-O hybridization gap due to the enhanced electron correlation.
This paramagnetic insulating state also exhibit a multi-configurational mixture of both HS and LS states with strong spin and charge fluctuations.
This DMFT result is different from the static DFT+U result showing various meta-stable solutions.  
The MS state in the BD structure accompanies the HS state in a LB Co site and the LS state in a SB site.
Charge ordering is also induced from this spin-state ordering since HS favors the $d^6$ charge state while LS occupies more $d^7$ states as covalently bonded with O ions.
The LB site with HS becomes a Mott insulator while the SB site with LS behaves as a band insulator.
Our DFT+DMFT calculation reveals that energetics between homogeneous and MS states are very close while DFT+U energetics depend sensitively on spin-states and structures favoring the MS state.

We also find that the Co-O covalency plays a crucial role in electronic structure of LaCoO$_3$.
Changing the double-counting potential in DFT+DMFT can lead to a different $N_d$ value tuning the covalency effectively and both energetics and one-particle spectra are more consistent with experiments when $N_d\sim6.7$. 
Increasing the Co-O covalency produces the insulator-to-metal transition faving the LS state.
Increasing temperature beyond 500K also reduces the correlation effect and drives the insulator-to-metal transition, similarly as the experiment.
The charge-self-consistency effect within DFT+DMFT reduces the tendency toward spin-state and charge orderings in LaCoO$_3$ producing close energetics between different structures compared to the non-charge-self-consistent calculation.
%Therefore, DFT+DMFT can be a promising method for describing the strong interplay among spin, charge, and lattice degrees of freedom in transition metal oxides.

% The volume expansion due to the spin-state transition produces a uniform spin-state excitation with a mixture of both HS ($S_z$=1.5,2.0) and LS ($S_z$=0,0.5) states fluctuating dynamically while the LS volume favors mostly the LS state.
%The structural distortion of the Co-O BD converges to a MS state with HS in the LB site and LS in the SB site.
%This spin-state ordering also accompanies the charge ordering since the LB site with HS weakens the Co-O covalency and reduces $N_d$ than the more covalently bonded SB site.
%Both expanded-volume structures are energetically almost degenerate and favored than the LS structure.
%The Co-O covalency plays an important role in explaining the strong spin and charge fluctuations occurring in LaCoO$_3$.
%We find that the conventional DFT+U method can converge to multiple static spin-state solutions which are energetically stable and their energetics are  sensitively dependent on the nature of spin-states and the Co-O covalency effect while DFT+DMFT does not suffer from these problems as the dynamical fluctuation effect is allowed.
%with the strong interplay among spin, charge, and lattice degrees of freedom.

\section*{Acknowledgments}
H. Park acknowledges helpful discussions with Andrew Millis.
H. Park and R. Nanguneri are supported by the U.S. Deptartment of Energy (DOE), Office of Science, Basic Energy Sciences (BES), Materials Sciences and Engineering Division.
A. Ngo is supported by the Vehicle Technologies Office (VTO), Department of Energy (DOE), USA.
We also gratefully acknowledge the computing resources provided on Bebop, a high-performance computing cluster operated by the Laboratory Computing Resource Center at Argonne National Laboratory. 

%\bibliography{main}

%\newpage

\section*{Appendix A: The $d-$orbital Hamiltonian}

%$$

In this Appendix, we show the matrix elements of the multi-orbital Hamiltonian, $\hat{H}^{dd}$, in the Co 3$d$ shell at each correlated site for three structures (S1, S2, S3 LB, and S3 SB) used in the DFT+DMFT calculations of this paper. 
First, we provide $\hat{H}^{dd}$ (in units of eV) represented using MLWF $d-$orbitals obtained from the DFT solution, which is relevant to the non-charge-self-consistent DFT+DMFT calculation.
The orbital order representing $\hat{H}^{dd}$ is $d^{3z^2-r^2}$, $d^{xz}$, $d^{yz}$, $d^{x^2-y^2}$, and $d^{xy}$.
% for different structures across the spin-state transition.
%First, we present the $\hat{H}^{dd}$ elements (in units of eV) computed within DFT (without charge updates from DMFT).

\begin{equation}
\hat{H}^{dd}_{S1}=
\begin{pmatrix}
  6.912 &  0.006 & -0.05  & -0.00  & -0.01  \\   
  0.006 &  6.236 &  0.019 & -0.01  & -0.02  \\
 -0.05  &  0.019 &  6.241 & -0.02  & -0.02  \\
 -0.00  & -0.01  & -0.02  &  6.917 & -0.01  \\
 -0.01  & -0.02  & -0.02  & -0.01  &  6.237 \\
\end{pmatrix}
\end{equation}

\begin{equation}
\hat{H}^{dd}_{S2}=
\begin{pmatrix}
  6.575  & -0.01   & -0.050 & -0.00   & 0.009 \\  
 -0.01   &  5.948  & -0.02  &  0.013  &-0.02  \\
 -0.050  & -0.02   &  5.954 & -0.02   & 0.020 \\
 -0.00   &  0.013  & -0.02  &  6.579  & 0.010 \\
  0.009  & -0.02   &  0.020 &  0.010  & 5.949 \\
\end{pmatrix}
\end{equation}

\begin{equation}
\hat{H}^{dd}_{S3,LB}=
\begin{pmatrix}
  6.484  &  0.003 &   0.054 &  -0.00  &   0.005 \\  
  0.003  &  5.913 &  -0.02  &  -0.00  &   0.018 \\
  0.054  & -0.02  &   5.921 &   0.027 &  -0.02  \\
 -0.00   & -0.00  &   0.027 &   6.489 &   0.002 \\
  0.005  &  0.018 &  -0.02  &   0.002 &   5.915 \\
\end{pmatrix}
\end{equation}

\begin{equation}
\hat{H}^{dd}_{S3,SB}=
\begin{pmatrix}
  6.583  &  0.006  &  0.049  & -0.00  &   0.014 \\  
  0.006  &  5.909  & -0.02   & -0.02  &   0.022 \\
  0.049  & -0.02   &  5.914  &  0.013 &  -0.02  \\
 -0.00   & -0.02   &  0.013  &  6.587 &   0.012 \\
  0.014  &  0.022  & -0.02   &  0.012 &   5.910 \\
\end{pmatrix}
\end{equation}

First of all, the local axis for initial projections of MLWFs are chosen to be aligned to the Co-O octahedron axis, therefore the off-diagonal elements are close to zeros.
The crystal-field splittings between $e_g$ and $t_{2g}$ orbitals are $\sim 0.67$eV for S1, $\sim 0.63$eV for S2, $\sim 0.57$eV for S3 LB, and $\sim 0.68$eV for S3 SB. 
As expected, the Co-O bond length can tune the crystal-field splitting (the energy difference between $e_g$ and $t_{2g}$ orbitals). 
Namely, the long-bond Co ion has the smaller splitting and favors higher spin-states.      
%obtained as the energy difference between $e_g$ and $t_{2g}$ orbitals. The LS structure $\Delta$ is $\sim 0.67$eV and 

The charge-self-consistency effect in DFT+DMFT produces a new charge density $\rho$ which is different from the original DFT $\rho$. As a result, $\hat{H}^{dd}$ is computed from the updated $\rho$ and is also changed. Here, we show below $\hat{H}^{dd}$ obtained from charge-self-consistent DFT+DMFT calculations using U=6eV, J=0.9eV, and the DC parameter $\alpha$=0.6.
In this case, the crystal-field splittings are $\sim 0.6$eV for S1 and S2, $\sim 0.66$eV for S3 LB, and $\sim 0.54$eV for S3 SB.
The main effect of the charge-self-consistency on $\hat{H}^{dd}$ is to reduce the crystal-field splittings for S1 and S2 structures compared to the non-charge-self-consistency case, therefore it promotes the homogeneous spin-state transition.
However, the charge-self-consistency effect also compensates for the spin-state and charge orderings as the crystal-field splitting has been increased for the LB site while it is decreased for the SB site in the S3 structure.
The average $d-$orbital level for the LB site becomes also much lower than the SB site, again compensating for charge ordering.

\begin{equation}
\hat{H}^{dd}_{S1}=
\begin{pmatrix}
  4.332 &  -0.01  &  -0.01  &  -0.00  &  0.014 \\  
 -0.01  &   3.737 &  -0.01  &   0.022 &  0.006 \\
 -0.01  &  -0.01  &   3.737 &  -0.02  &  0.004 \\
 -0.00  &   0.022 &  -0.02  &   4.333 &  0.022 \\
  0.014 &   0.006 &   0.004 &   0.022 &  3.735 \\
\end{pmatrix}
\end{equation}
\begin{equation}
\hat{H}^{dd}_{S2}=
\begin{pmatrix}
  3.767 &  0.013 & -0.01 & -0.00 & -0.02 \\  
  0.013 &  3.158 &  0.006 & -0.02 &  0.007 \\
 -0.01 &  0.006 &  3.157 & -0.02 & -0.01 \\
 -0.00 & -0.02 & -0.02 &  3.767 & -0.02 \\
 -0.02 &  0.007 & -0.01 & -0.02 &  3.156 \\
\end{pmatrix}
\end{equation}
%$$

\begin{equation}
\hat{H}^{dd}_{S3,LB}=
\begin{pmatrix}
  3.150 &  -0.02  &  0.016 &  -0.00  &  -0.02  \\ 
 -0.02  &   2.490 &  0.009 &   0.037 &  -0.01  \\
  0.02  &   0.009 &  2.488 &   0.033 &   0.006 \\
 -0.00  &   0.037 &  0.033 &   3.148 &  -0.04  \\
 -0.02  &  -0.01  &  0.006 &  -0.04  &   2.487 \\
\end{pmatrix}
\end{equation}

\begin{equation}
\hat{H}^{dd}_{S3,SB}=
\begin{pmatrix}
  4.519 & -0.01  &  0.006 &  -0.00  &  -0.01  \\  
 -0.01  &  3.984 &  0.006 &   0.017 &  -0.01  \\
  0.006 &  0.006 &  3.984 &   0.014 &   0.005 \\
 -0.00  &  0.017 &  0.014 &   4.520 &  -0.02  \\
 -0.01  & -0.01  &  0.005 &  -0.02  &   3.983 \\
\end{pmatrix}
\end{equation}

\section*{Appendix B: Double counting correction}

A frequently used expression of the double-counting energy $E^{DC}$ is the fully localized limit (FLL) form~\cite{PhysRevB.44.943} which has been adopted frequently in DFT+U.
\begin{eqnarray}
E^{DC}=\frac{U}{2}\cdot N_d\cdot(N_d-1)-\frac{J}{4}\cdot N_d\cdot(N_d-2)\\
\label{eq:Edc0}
V^{DC}=\frac{\partial E^{DC}}{\partial N_d}=U\cdot(N_d-\frac{1}{2})-\frac{J}{2}\cdot(N_d-1)
\label{eq:Vdc0}
\end{eqnarray}
where $N_d$ is the occupancy of the correlated site and can be obtained as the result of self-consistent DFT+DMFT or DFT+U calculations. 
Therefore, the $V^{DC}$ potential depends on the correlated site since $N_d$ is site-dependent.

Recently, it has been shown that the exact form of $V^{DC}$ within DFT+DMFT~\cite{PhysRevLett.115.196403} can be computed and the formulae should be close to the nominal DC form, where $N_d$ in Eq.$\:$\ref{eq:Vdc0} is replaced to the nominal $d-$occupancy in the atomic limit, $N_d^0$, which is site-independent.
\begin{eqnarray}
V^{DC}=U\cdot(N_d^0-\frac{1}{2})-\frac{J}{2}\cdot(N_d^0-1)
\label{eq:Vdc2}
\end{eqnarray}
The hybridization of $d$ and $p$ orbitals in transition metal oxides means that the resulting $d-$occupancy $N_d$ will be larger than the nominal value $N_d^0$ ($N_d>N_d^0$), therefore the nominal $V^{DC}$ will be always smaller than the FLL $V^{DC}$.
This smaller $V^{DC}$ potential reduces the covalency effect between $d$ and $p$ orbitals.

In this paper, we use the following forms of $E^{DC}$ and $V^{DC}$ by modifying the FLL forms to allow the tuning of $V^{DC}$ for changing the Co-O covalency effect.
\begin{eqnarray}
E^{DC}=\frac{U}{2}\cdot\overline{N_d}\cdot(\overline{N_d}-1)-\frac{J}{4}\cdot\overline{N_d}\cdot(\overline{N_d}-2)\\
\label{eq:Edc1}
V^{DC}=U\cdot(\overline{N_d}-\frac{1}{2})-\frac{J}{2}\cdot(\overline{N_d}-1)
\label{eq:Vdc1}
\end{eqnarray}
%where $\overline{N_d}=N_d-\alpha$. 
where $\overline{N_d}=N_d-\alpha$ with a parameter $\alpha$. 
Our $V^{DC}$ formula can be derived from $E^{DC}$ ($V^{DC}=\partial E^{DC}/\partial N_d$) and allow site-dependent potentials similarly as the FLL form.   
The conventional FLL DC form is recovered by setting $\alpha$=0. 
By increasing $\alpha$, $V^{DC}$ can be close to the nominal $V^{DC}$ value as $\overline{N_d}$ approaches to $N_d^0$ ($\alpha=N_d-N_d^0$). 

Another modified form of $V^{DC}$ for DFT+DMFT was also suggested as below since the $U$ value used in the FLL form can be smaller than the Hubbard $U$ ($U^{\prime}<U$) to allow the smaller $V^{DC}$ potential than the FLL one:
\begin{eqnarray}
E^{DC}=\frac{U^{\prime}}{2}\cdot N_d\cdot(N_d-1)-\frac{J}{4}\cdot N_d\cdot(N_d-2)\\
V^{DC}=U^{\prime}\cdot(N_d-\frac{1}{2})-\frac{J}{2}\cdot(N_d-1)
\label{eq:Vdc3}
\end{eqnarray}
where $U^{\prime}=U-\alpha$ with a parameter $\alpha$.
Here, the role of $\alpha$ is the same as the one in Eq.$\:$\ref{eq:Vdc1}. 
Namely, the covalency effect can be reduced by increasing the $\alpha$ value. 
It has been shown that using $U^{\prime}=U-0.2$eV can successfully reproduce the structural and electronic phase diagram of rare-earth nickelates~\cite{PhysRevB.89.245133,PhysRevB.90.235103}.

\bibliography{main}

\end{document}